Exploiting deep learning in forecasting the occurrence of severe haze in Southeast Asia

Chien Wang
Laboratoire d'Aerologie, CNRS/UPS, Toulouse, France
chien.wang@aero.obs-mip.fr or wangc@mit.edu

**Abstract**
Severe haze or low visibility event caused by particulate pollution has become a serious environmental issue in Southeast Asia. A forecasting framework of such events based on deep convolutional neural networks has been developed. The framework has been trained using time sequential maps of up to 18 meteorological and hydrological variables alongside surface visibility data over past 35 years. In forecasting haze versus no-haze situations in Singapore, the trained machine has achieved a good overall accuracy that easily exceeds that of the no-skill "blinded forecast" based on haze occurrence frequency. However, the machine still produces a relatively high number of missing forecasts (false negative for haze events), likely owing to its lack of "experience" in identifying atypical patterns. Nevertheless, this effort has demonstrated a promising prospect of using deep learning algorithms to predict the occurrence of extreme environmental and weather events, and to advance knowledge about these still poorly known phenomena.

**1. Introduction**

Adverse environmental and weather events can cause substantial economic loss. For example, particulate pollution has become a serious environmental issue in many Southeast Asian countries in recent decades, where elevated aerosol abundance frequently cause low visibility situations or haze events to occur, threatening human health and interrupting working, outdoor, and school activities as well as transportations (*e.g.*, Lee *et al*., 2017). It has been identified that planned and unplanned fires in Southeast Asia are major reasons behind the severe while often short-lasting haze events (*i.e.*, with a daily mean visibility, or *vis.* < 7 km) in the region, whereas non-fire anthropogenic activities are likely responsible for more frequent while less drastic haze events (*i.e.*, with 7 km < *vis.* < 10 km) (Lee *et al*, 2017, 2018, 2019).

To ultimately prevent such severe hazes to occur requires rigid emission control measures in place, through significant changes in energy consumption as well as land and plantation management, dealing with economic planning, legislation, trade negotiation, among others, in all involved countries. Before the above measures could finally take place, for reducing economic loss it would actually be more practical to develop skills to accurately predict the occurrence of hazes and hence to allow certain mitigation measures to be implemented ahead of time.

Various types of deterministic numerical models developed based on fluid dynamics and explicit or parameterized representations of physical and chemical as well as radiative processes, ranging from global, regional, to local scales, are supposed to be the tools for predicting the occurrence of adverse environmental or weather extremes such as hazes in Southeast Asia. However, this is still a difficult task for reasons. First of all, the complex multidisciplinary nature required for such a task might have already exceeded the scopes of many of these bottom-up-designed or process-orientated models. More critically, low-probability extreme events are associated with significant and sophisticated spatiotemporal variabilities, and often occur abruptly. The propagation of numerical or parameterization errors throughout forward integration of the nonlinear models often cloud the forecasting signals. On the other hand, lack of knowledge about the extreme events, in turn, also hinders the further improvement of forecasting skills. For instance, even after adopting expensive measures such as data assimilation (Kalnay, 2002) and ensemble simulations (Lewis, 2005), forecast of, *e.g.*, storm track made by using weather models could still miss the target because *a posteriori* distribution of the possible outcomes of the targeted extreme event might not perfectly overlap with that reflected from the results of these models. Specifically, for forecasting haze events, the process-orientated models need to accurately predict the concentration of aerosols at a given geographic location and time in order to correctly derive surface visibility. This is a task that relies on the numerical representations of processes ranging from emissions, atmospheric transport, mixing, to deposition of particulate matters, in addition to meteorological conditions. Lack of real-time emission data alone would simply handicap such an attempt.

On the other hand, technological advancement and continuous investment from governments across the world have led to a rapid increase of quantity alongside substantially improved quality of meteorological, oceanic, hydrological, land, and atmospheric composition data, providing resources for developing "task-oriented" forecasting models without involving a deterministic framework with deeply connected process layers. These data might still not be sufficient for evaluating and improving certain detailed aspects of the traditional forecast models. Nevertheless, they perhaps have already contained rich information of environmental conditions favoring the occurrence of the extreme events of interest and thus are of great values to the effort of developing forecasting skills based on machine learning or deep learning algorithms (*e.g.*, LeCun *et al.*, 2015).

Southeast Asian haze events have been previously studied by using sophisticated atmospheric models to simulate the atmospheric evolution of anthropogenic aerosols during selected time periods in the past (*e.g.*, Lee *et al.*, 2017, 2018). The modeled aerosol concentration is found to be not accurate enough to reproduce measured surface visibility at given measurement sites in the region including Singapore. Only after corrections being made based on in-situ aerosol measurements could the model manage to reproduce about 80% haze events (defined as events with daily surface visibility lower than 10 km; Lee *et al.*, 2018), or an accuracy equivalent to the "training accuracy" in machine learning. In addition, several machine learning algorithms including logistic regression, random forest, support vector machine, and multilayer perceptron, have then been applied to perform experimental forecasting of haze events for several sites in Southeast Asia, using certain abstract features derived based on human opinion as inputs. The results are generally around about 90% training accuracy using data for the same-day of haze and 84% using data with 1-day leading time (*i.e.* previous day) (Lee *et al.*, 2018). Further improvement is difficult due to limited knowledge about this phenomenon.

Deep learning (or DL), on the other hand, appears to have certain advantage for the task comparing to the commonly applied machine learning algorithms. It directly links large quantity of raw data with targeted outcomes through deep neural networks or sophisticated correlations (Goodfellow *et al.*, 2016), and thus can actually avoid possible mistakes in data derivation or selection introduced by subjective human opinion regarding a poorly understood phenomenon. Recently, deep learning algorithms have been explored for various applications in atmospheric, climate, and environmental sciences, ranging from recognizing specific weather patterns (*e.g.*, Liu *et al.*, 2016; Kurth *et al.*, 2018; Lagerquist *et al.*, 2019; Chattopadhyay *et al.*, 2020), weather forecasting including hailstorm detection (*e.g.*, Grover *et al.*, 2015; Shi *et al.*, 2015; Gagne *et al.*, 2019), to deriving model parameterizations (*e.g.*, Jiang *et al.*, 2018), and perhaps beyond. Here presented is a framework for forecasting the occurrence of severe haze events in Southeast Asia, specifically centered at Singapore. It has been developed using deep convolutional neural networks or CNNs and trained using a large quantity of meteorological and hydrological data. The methodology is firstly described after the Introduction, followed by presentations and discussions of training and validation results and then a summary.

## 2. Methodology

The CNN-based framework, the HazeNet, has been developed by largely following the architecture of the CNN developed by the Oxford University's Visual Geometry Group or VGG-Net (Simonyan and Zisserman, 2015). The Python written script uses Keras interface library (https://github.com/keras-team/keras) with TensorFlow-GPU (https://www.tensorflow.org) as backend. The actual architectures alongside hyper-parameters have been selected based on numerous training tests. In addition, certain techniques such as batch normalization (Ioffe and Szegedy, 2015) that was not available when the original VGG net was developed, have been adopted as well. The architecture of a 16-layer machine can be briefly described as in Table 1.

**Table 1**. The architecture of HazeNet-16 (H-16)

| Layers | Sizes or filter Sets | Kernel |
|---|---|---|
| Input | 40x40x18 or 60x60x18 | |
| conv 1 + dropout; conv 2 | 92, 92 | 10x10, 10x10 |
| Maxpool | | 2x2 |
| conv 3 + dropout; conv 4 | 192, 192 | 6x6, 6x6 |
| Maxpool | | 2x2 |
| conv 5 + dropout; conv 6 | 384, 384 | 3x3, 3x3 |
| Maxpool | | 2x2 |
| conv 7 + dropout; conv 8 | 384, 384 | 3x3, 3x3 |
| Maxpool | | 2x2 |
| conv 9 + dropout; conv 10 | 512, 512 | 3x3 (or 1x1), 3x3 (or 1x1) |
| Maxpool | | 2x2 |
| conv 11 + dropout; conv 12 | 512, 512 | 3x3 (or 1x1), 3x3 (or 1x1) |
| Maxpool | | 2x2 |
| Flatten | | |
| dense1, dense2 + dropout | 4096, 4096 | |
| sigmoid/softmax | 1, 2 or 3… | |

**Note**: Here conv = ZeroPadding layer + Convolution layer + BatchNormalization layer. H-16 has 35,407,417 parameters including 24,688 non-trainable ones.

The network has been trained in a standard supervised learning procedure for classification. The label for classes is defined using the observed daily surface visibility (*vis.* thereafter) in Singapore from 1982 to 2016, obtained from the Global Surface Summary Of the Day or GSOD dataset consisting of daily observations of meteorological conditions from tens of thousands of airports around the globe (Smith *et al.*, 2011). The total sample number is 12,784. The same dataset with a slightly shorter coverage was used in our previous attempt with various machine learning algorithms (Lee *et al.*, 2018). Differing from the previous study too, here the classes are defined using given percentiles of observed *vis.* during 1982-2016, *i.e.*, the $5^{th}$ percentile (or p5) of 7.72 km, $10^{th}$ percentile (or p10) of 8.85 km, $25^{th}$ percentile (or p25) of 9.98 km, or the medium of 10.78 km. For example, in a 2-class training, events with *vis.* $\leq$ p5 are defined as class 1, otherwise the class 0; and in a 3-class training, class 2 = *vis.* $\leq$ p5, class 1 = p5 < *vis.* $\leq$ p25, class 0 = otherwise. Note that fog is a low visibility event too. However, there have been only 18 fog days recorded in Singapore during 1982 and 2016, not mentioning that fog formation might also relate to aerosol abundance, therefore, they are not specifically identified in labeling.

The input data for training HazeNet are longitude-latitude maps of meteorological and hydrological variables covering the Southeast Asia domain (Fig.1, left panel), representing different aspects of weather from status to circulations, obtained from the reanalysis dataset of ERA-Interim produced by the European Centre for Medium-range Weather Forecasts or ECMWF (Dee *et al*., 2011). These daily data are distributed in a grid system with spatial intervals of 0.75 by 0.75 degree or approximately 80 km. Up to 18 variables are used, including surface temperature (T1000), relative humidity (RelHum), horizontal wind speeds at 10-meter above the surface (U10 and V10), total water and water vapor in the atmospheric column (TCW and TCWV), convective and large-scale precipitation (ConvPrecip and LgScalePrecip), height of planetary boundary layer (BLH), geopotential heights at 850 and 500 hPa pressure levels (Z850 and Z500), total, high, middle, and low cloud covers (Tcloud, HCloud, MCloud, and LCloud), and water volume in 3 different soil layers (SWVL1, SWVL2, and SWVL3) (see Fig.1 right panel for the examples of inputs for August 10, 2082 – a severe haze day with *vis*. = 7.56 km). The inputs have been provided to the CNN in a format as time sequence samples of three-dimensional fields of, *e.g*., 40x40x18 or 60x60x18.

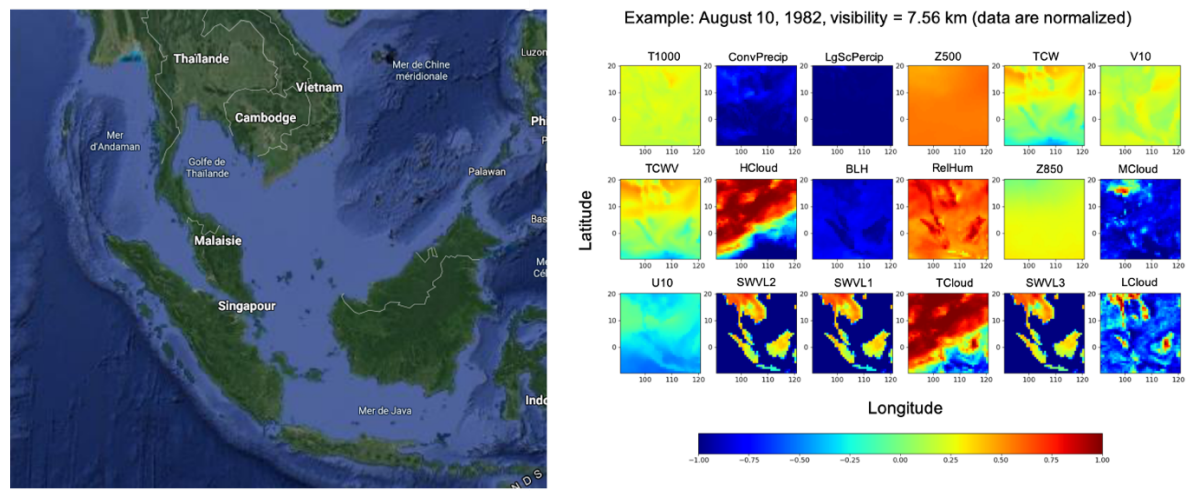

**Figure 1**. (Left) Geographical coverage of input data for 60x60 grids; (Right) Example of input features of August 10, 1982 or a severe haze day in Singapore with *vis*. = 7.56 km. Note that data are normalized into [-1, +1] using the entire time series of each given feature.

The principal of the approach adopted in this study is to apply the common practice in visual classification with deep CNN for the task, *i.e*., to train the CNN to identify weather and hydrological patterns associated with hazes and hence to advance our knowledge about favorite environmental conditions of such events. Obviously, the sample number is still somewhat limited for training a deep CNN as HazeNet. Nevertheless, the required data analysis dealing with 18 joint variables in a 35-year time period would already post a challenge to any methodology other than adopted deep learning framework.

The targeted severe hazes are low probability events. For p5, the frequency of its appearance is about 5.31% (679 in 12784 event-wise). Therefore, trained machine would easily bias toward recognizing the overwhelming non-haze events (with a 94.7% frequency of appearance). To overcome this issue, a class-weight or batch normalization, or a combination of both could be a solution as demonstrated in Fig. 2. Note that an additional benefit of using batch normalization is to make the training of a deep CNN become possible.

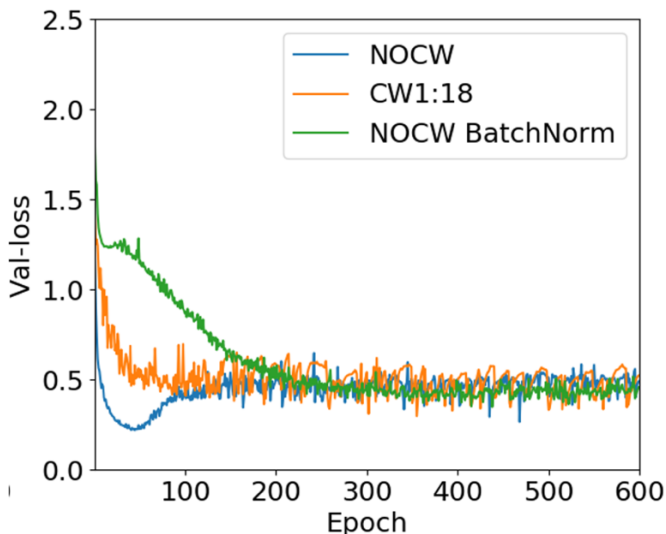

**Figure 2**. Validation losses of HazeNet when excluding class-weight (NOCW), including a class-weight in 1:18 (CW1:18), and including only batch normalization (BatchNorm).

## 3. Results

For simplicity, discussions will focus on H-16 2-class classifications. Trained H-16 can easily reach 95.2% or higher in validation accuracy for a 2-class classification case. Most trainings take 1000 epochs (in some cases 1500 epochs) to accomplish, mainly for achieving lower validation loss or error. Specifically, in forecasting targeted class 1 events, *i.e.*, severe hazes, the H-16 with 18 inputs can reach a precision of nearly 0.60, F1 score of 0.37, and all the other metrics are also quite reasonable (Fig. 3, left panel), especially for a task that lacks human forecasting skill to compare. Nevertheless, the current machine still produces a relatively large number of missing forecasts for class 1 events (*i.e.*, false negative; see Fig. 3, right panel).

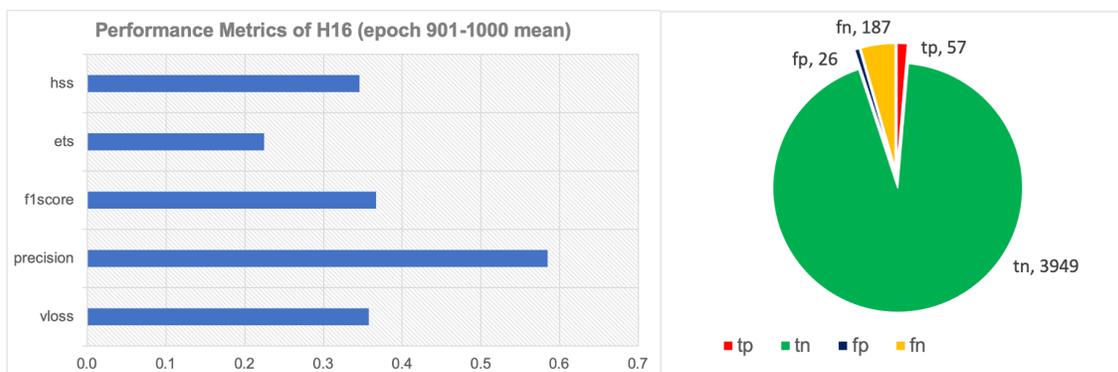

**Figure 3**. (Left) Performance metrics of H-16 in forecasting class 1 events for a 2-class clarification. (Right) Predicted outcomes categorized based on observations, tp, tn, fp, and fn = true positive, true negative, false positive, and false negative, respectively, all related to class 1 (severe haze). All data shown are validation results averaged over epoch 901-1000. Here *vloss* = validation loss; *precision* = $tp/(tp + fp)$; with *recall* = $tp/(tp + fn)$, *f1 score* = $2 \times precision \times recall/(precision + recall)$; equitable threat score *ets* = $(tp - tpr)/(tp + fn + fp - tpr)$; and Heidke skill score *hss* = $2 \times (tp \times tn - fp \times fn)/[(tp + fp) \times (fp + tn) + (tp + fn) \times (fn + tn)]$ with $tpr = (tp + fn) \times (tp + fp)/N$, and N = total sample number.

One interesting finding is that the actual feature patterns of the same predicted outcome judged by observation (truth), *i.e.*, true positive (correctly predicted severe haze), true negative (correctly predicted no-haze), false positive (false alarm), or false negative (missing forecast) differ quite significantly among themselves (Fig. 4, an example of true positive of SWVL1). Note that the difference among true positive outcomes shown in Fig. 4 only reflects the case of 1 out of 18 features. This demonstrates how difficult could be for a traditional case analysis, which often concentrates in a few selected cases partially due to the massive labors by human, to reach a generalized conclusion about the extreme events such as haze. Nevertheless, such results also suggest the necessity of developing additional tools to further reveal co-variability among various features.

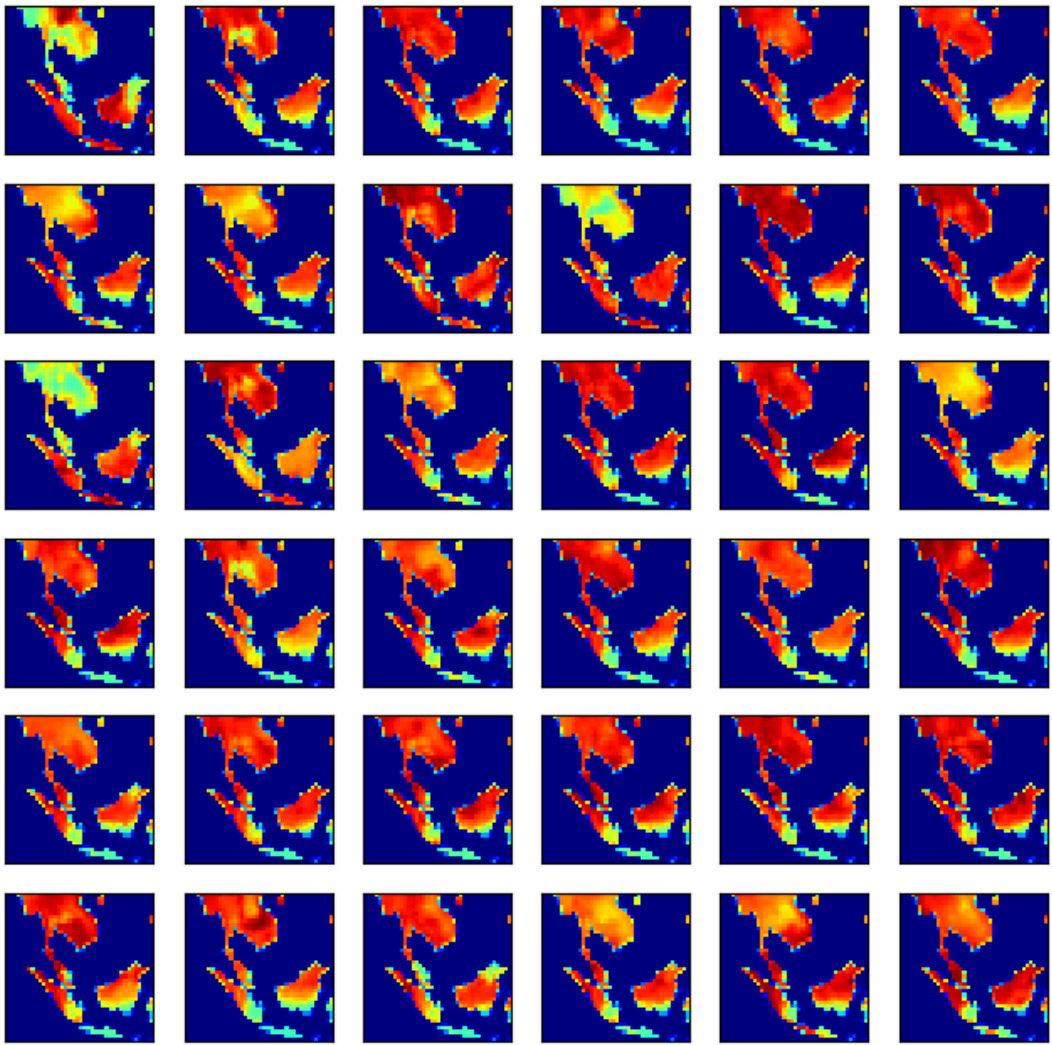

**Figure 4**. Patterns of soil water volume in the first soil layer (SWVL1) of the true positive outcome.

Although the feature patterns even within the same predicted outcome differ significantly, the averaged feature patterns of each of these outcomes do exhibit their own distinguishable characteristics (Fig. 5). In addition, looking into the averaged patterns of different outcome groups, one can also understand why the machine could mistakenly classify the events in a general sense. For instance, for the feature of soil water volume in the first layer, *i.e.*, SWVL1, the average pattern of false positive shares many characteristics with that of true positive except in some details mostly near the northern edge of the domain, while average pattern of false negative shares characteristics with true negative across most regions of Java island and Malay Peninsula (lower part of the domain). The similar situation occurs in the cases of TCWV as well. In-depth analysis on these details might provide useful lead of how to derive data maps, and to train the machine to weight more on the "useful" characteristics of patterns in order to correctly classify the outcomes. It is worth indicating that the actual pattern of a given event is consisting of up to 18 individual patterns of different variables. Therefore, only a joint analysis involving all patterns of various variables could serve the best purpose.

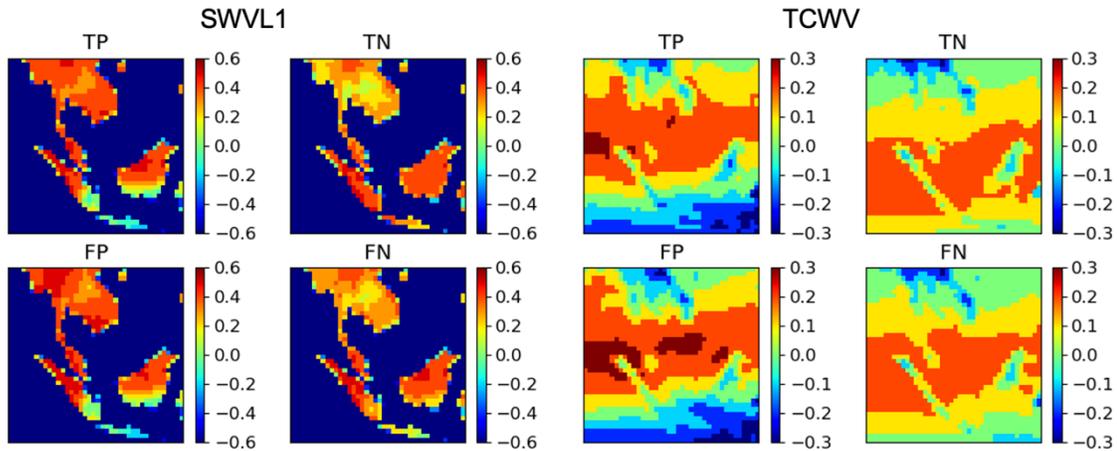

**Figure 5**. Average patterns of various outcomes from true positive, true negative, to false positive and false negative for: (Left) soil water valume in the first soil layer (SWVL1) and (Right) column water vapor or TCWV. The patterns are arithmetic means throughout all cases of each outcome category in validation. They were then blurred to emphasize their major characteristics by an operator of ROUND(value*10.0)/10.0, here value represents the value of a given mean pattern.

As a comparison, several deep CNN architectures have also been trained in the study alongside HazeNet, these include Inception Net or GoogLeNet (Version 3; Szegedy *et al.*, 2015), ResNet (ResNet50; He *et al.*, 2015), and VGG Net (VGG-19; Simonyan and Zisserman, 2015). These networks have been trained firstly in their original format, then in revised format to adopt certain hyper-parameters of HazeNet architecture. Judged by the metrics of validation including validation loss, F1 Score, Heidke skill score, and equitable threat score – here a higher score represents a better performance except for validation loss – H-16 appears to be the best overall performer in comparison (Fig. 6). For the other networks, in many cases the ones with revised architectures using HazeNet hyper-parameters performed better than their original one did. This indicates that in an application that uses meteorological data, it is important to design a network architecture that is suitable for recognizing characteristic meteorological patterns with unique spatial scales and features. Note that the original VGG-19 does not include batch normalization and cannot finish the training unless batch normalization is adopted. As a result, its scores are not included in Fig. 5. In terms of training cost, both InceptionV3 (21,708,961) and ResNet50 (23,614,465) have fewer number of parameters but require longer training times (~4 and 3 times, respectively, using Nvidia V100 GPUs) than HazeNet for the targeted purpose. While VGG-19 with batch normalization requires the similar training time as HazeNet does but has slightly more parameters (45,265,921 versus 35,407,417).

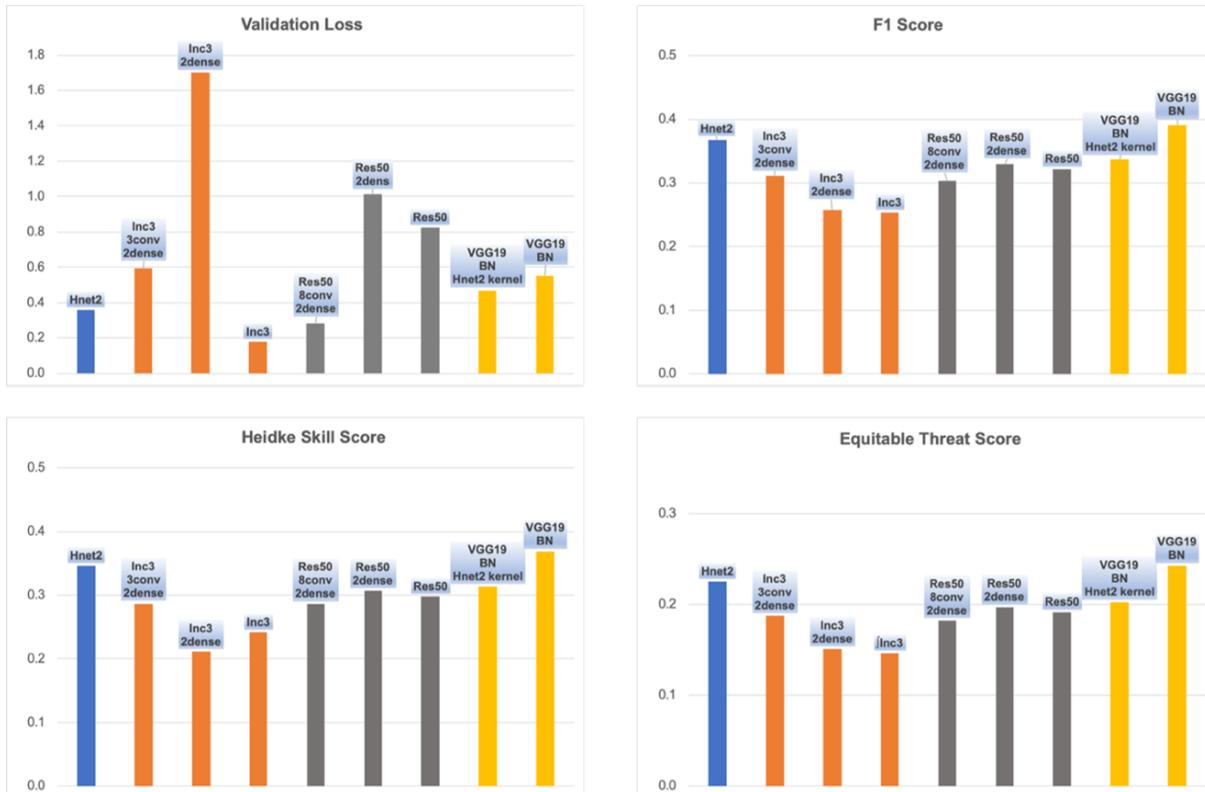

**Figure 5**. Comparison of validation metrics of networks with different architectures in a 2-class classification. Here Inc3 = inception v3, Res50 = Resnet50; Inc3-3conv-2dense = first 3 conv of H-16 + Inc3 + 2 dense of H-16; Res50-8Conv-2Dense = first 8 conv of H-16 + Res50 + 2 dense of H-16.

An important aspect of the trained machine is whether it has bias in predicting events in certain seasons or years. Since all the trainings started from a randomly shuffled dataset, for knowing the exact date of a given event it is more convenient to use the entire original datasets to conduct the above evaluation (not a true validation though). It shows that the trained H-16 does not display bias towards any specific year in terms of percentage of wrong forecasts. On the other hand, February and March (typically in non-fire season) are found to be the months when the machine makes more mistakes relatively, suggesting that the occurrence of severe hazes by non-fire aerosols perhaps is more difficult to predict than ones caused by fires (Fig. 7).

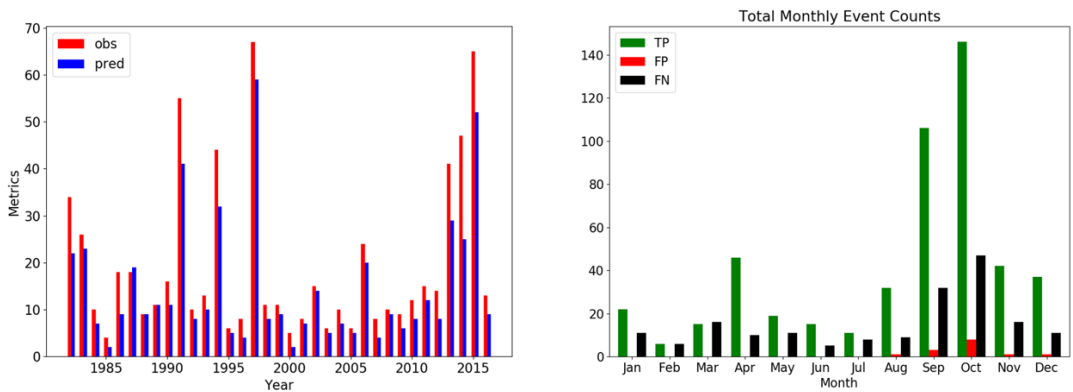

**Figure 7**. Comparison of predicted with observed severe haze events in different years (left) and the prediction outcomes in different months (right) based on evaluation using entire inputs.

## 4. Discussions

One anticipated advancement in this study is to identify the importance of various features to the machine in performing the forecasting task and thus to examine the favorite environmental conditions for hazes possibly with a fewer involved variables. The similar analysis for many machine learning algorithms ranging from regressions to decision trees is rather straightforward, *e.g.*, by using libraries such as scikit-learn (https://scikit-learn.org/). However, for deep CNNs, this task is rather difficult (*e.g.*, McGovern *et al.*, 2019). Here a simple method has been used for such a purpose. Firstly, two types of machines have been trained: (a) by excluding one different given input feature each time, hereafter Ext.1; and (b) by including only one input feature each time, or Ext.17. Then, the sensitivity of the CNN to a given feature is determined by its combining performance ranks from both sets. All the trainings are for the same 2-class classification with H-16.

The sensitivity of each feature is ranked by its significance in: (a) causing the performance degradation in Ext.1 or (b) avoiding performance degradation in Ext.17 in reference to the standard H-16, measured by a linear combination of several selected metrics as:

$$\text{score} = \Delta F1score + \Delta hss + \Delta ets - \Delta vloss \qquad (1)$$

Here $\Delta$ is the percentage change of a given metrics in reference to H-16 result; *hss*, *ets*, and *vloss* represents Heidke skill score, equitable threat score, and validation loss, respectively. For all the metrics except vloss, higher value means better performance. Due to the deep architecture of HazeNet, Ext.1 is much less sensitive to the selection of included feature than Ext.17 – some Ext.17 runs have shown signs of overfitting or lack of convergence (Fig. 6). Therefore, the overall rank is derived by weighting 30% on Ext.1 rank and 70% on Ext.17 rank (Table 2).

**Table 2. Relative sensitivity rankings of various input features in classification**

| Features | Ext.17 ranking | Ext.1 ranking | Overall ranking |
|---|---|---|---|
| SWVL1 | 1 | 4 | 1 |
| SWVL2 | 2 | 5 | 2 |
| TCWV | 6 | 1 | 3 |
| RelHum | 4 | 8 | 4 |
| T1000 | 5 | 11 | 5 |
| SWVL3 | 3 | 17 | 6 |
| U10 | 10 | 3 | 7 |
| V10 | 8 | 9 | 8 |
| TCW | 7 | 18 | 9 |
| MCloud | 14 | 2 | 10 |
| Z850 | 12 | 7 | 11 |
| BLH | 9 | 15 | 12 |
| Z500 | 15 | 6 | 13 |
| LCloud | 11 | 16 | 14 |
| TCloud | 13 | 13 | 15 |
| HCloud | 16 | 12 | 16 |
| LgScalePrecip | 18 | 10 | 17 |
| ConvPrecip | 17 | 14 | 18 |

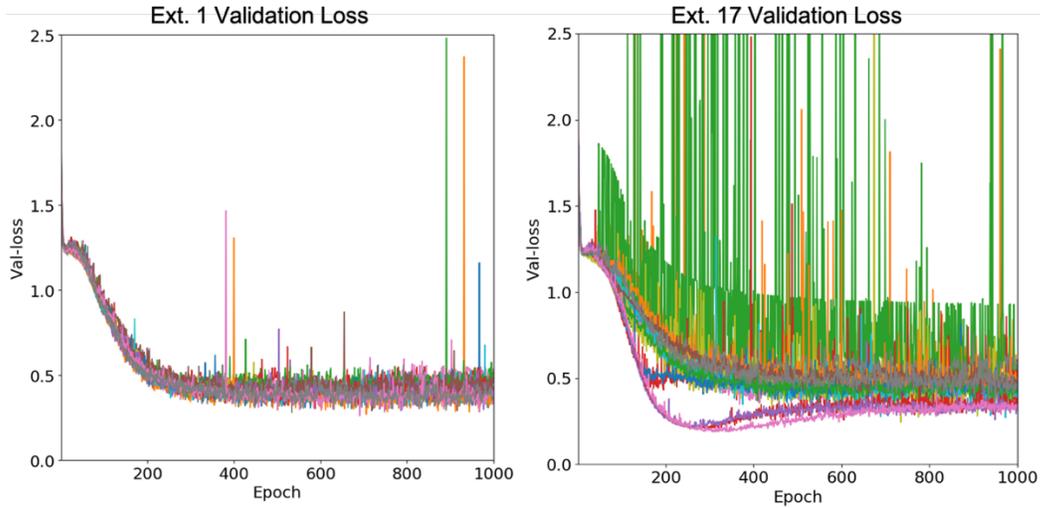

**Figure 6**. Validation loss (error) of various machines trained in Ext.1 (left panel) and Ext.17 (right panel).

The results of Ext.17 and Ext.1 suggest that soil water volumes alongside humidity measured by column water vapor content and relative humidity are among the important features in forecasting the occurrence of severe hazes in Singapore. Soil water volume in Southeast Asian islands determines the level of water table of many peatlands where peat burning could last as long as water table is below the top of peat layer (*e.g.*, Page and Hooijer, 2016). On the other hand, relative humidity and water vapor content are key factors in causing haze to occur – high relative humidity favors water condensation on the surface of aerosols and thus enhances light extinction of these aerosols (*e.g.*, Pruppacher and Klett, 2010). In an experiment to only include 4 top ranking features in Table 2, *i.e.*, SWVL1, SWVL2, TCWV, and RelHum, the trained machine has achieved a performance that is only slightly inferior to that of H-16 (Fig. 7), indicating the possibility of training a good-performing machine with fewer features. Despite that the performance dependency on multiple features are complicated due to the deep architecture of HazeNet, this is still an interesting direction for further efforts.

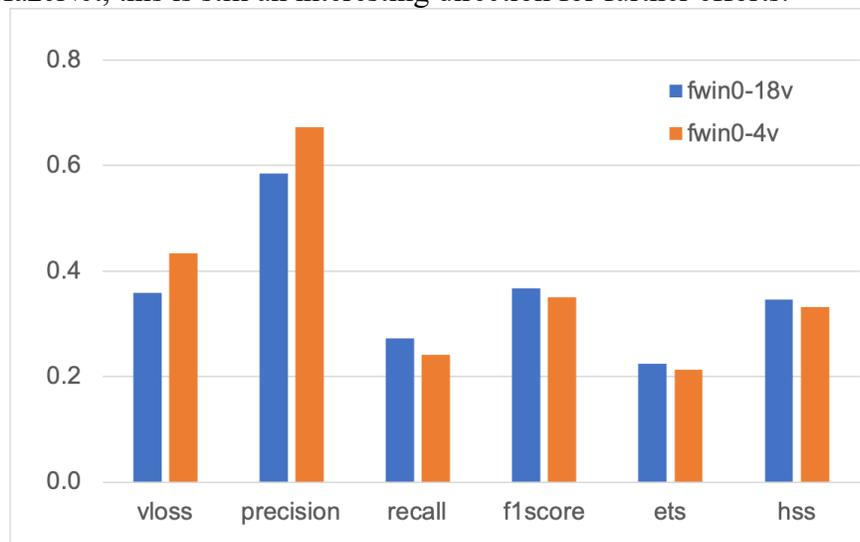

**Figure 7**. Comparison in major performance metrics between H-16 with 18 versus 4 input features. Here higher values represent better performance except for vloss.

For a forecasting task, the leading time (*i.e.*, time ahead of the occurrence of the event) of input data used in making prediction is always a critical aspect. Note that weather forecast today can provide at least one-week advance predictions and thus in theory one can always have the same day data in practice. Nevertheless, exploring this issue can actual help us to better understand the nature of a forecasting framework that does not rely on the deterministic correlations. As shown in Fig. 8, in a deterministic framework, forecasting the status of a given feature at a given time is constrained by the status at previous time steps through a theoretical framework or dynamical equation, *i.e.*, function *f* in the figure. When error rises at any of these steps, the forecasted status could drift away from the real one though following a rate suggested by the equation. On the other hand, a DL framework would only look for a predictor of the feature status at a given time (or an event corresponding to such a status), such a predictor can be a feature status at some given leading time or a combination of status at a few steps with different leading times. As far as a correlation is established between the predictor and the target, the forecast can be proceeded with the statistical confidence estimated from validation.

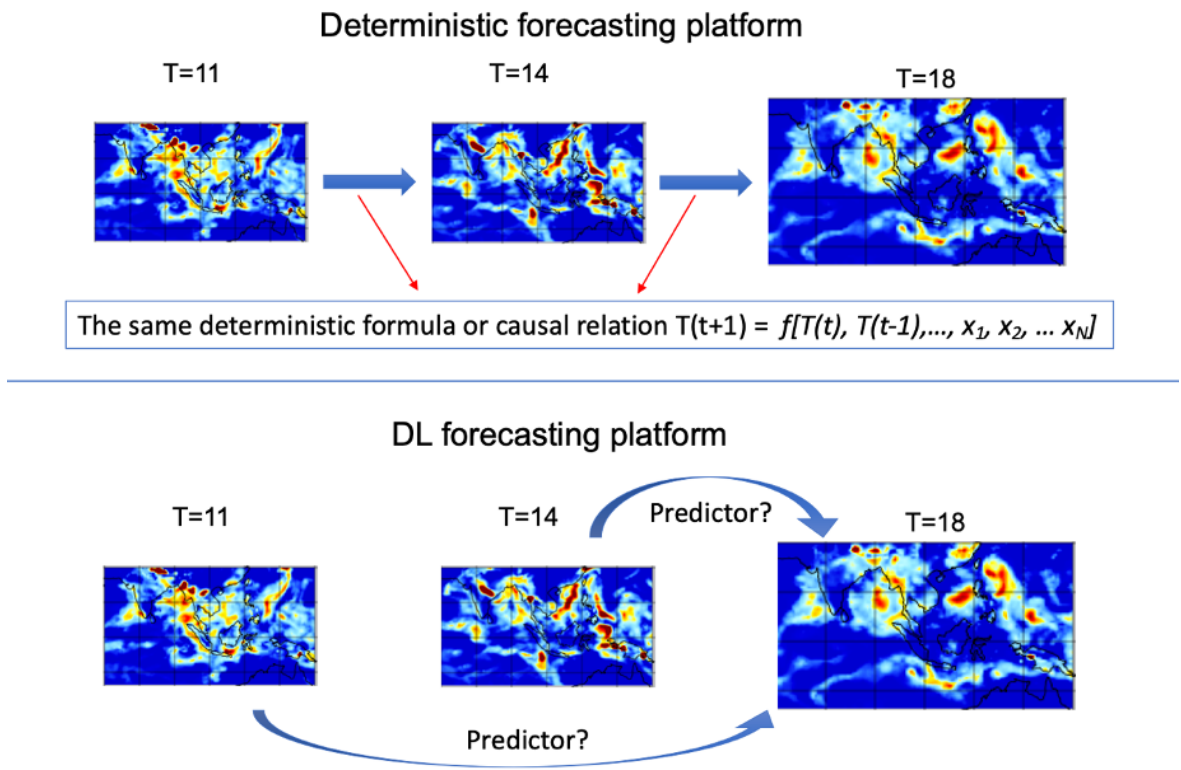

**Figure 8.** Diagram of the differences between a deterministic and a DL forecasting framework. Here T=11, 14, and 18 represents the status of a given feature at different times, *f* represents the dynamical equation that constrains the deterministic framework, and $x_j$ represents other parameters.

Interestingly, the above prospective is reflected in the experiment of training machines with input features at different leading times, here from the same day or Fwin = 0 to 5 days or Fwin = 5. As shown in Figure 9, the average patterns of true positive outcomes derived with various leading times appear to be quite different, indicating that the targeted status of the involved feature could correspond to different types of past feature status due to the shift of leading time.

Nevertheless, they could all be useful as predictors to forecast the occurrence of the same target event. Note that the patterns of true negative through different leading times differ insignificantly, largely due to the fact that this type of events in a dominate number overlap in the trainings of various machines. It is worth indicating though that the performance of the machines does slightly degrade specifically when leading time is about 4 days or longer.

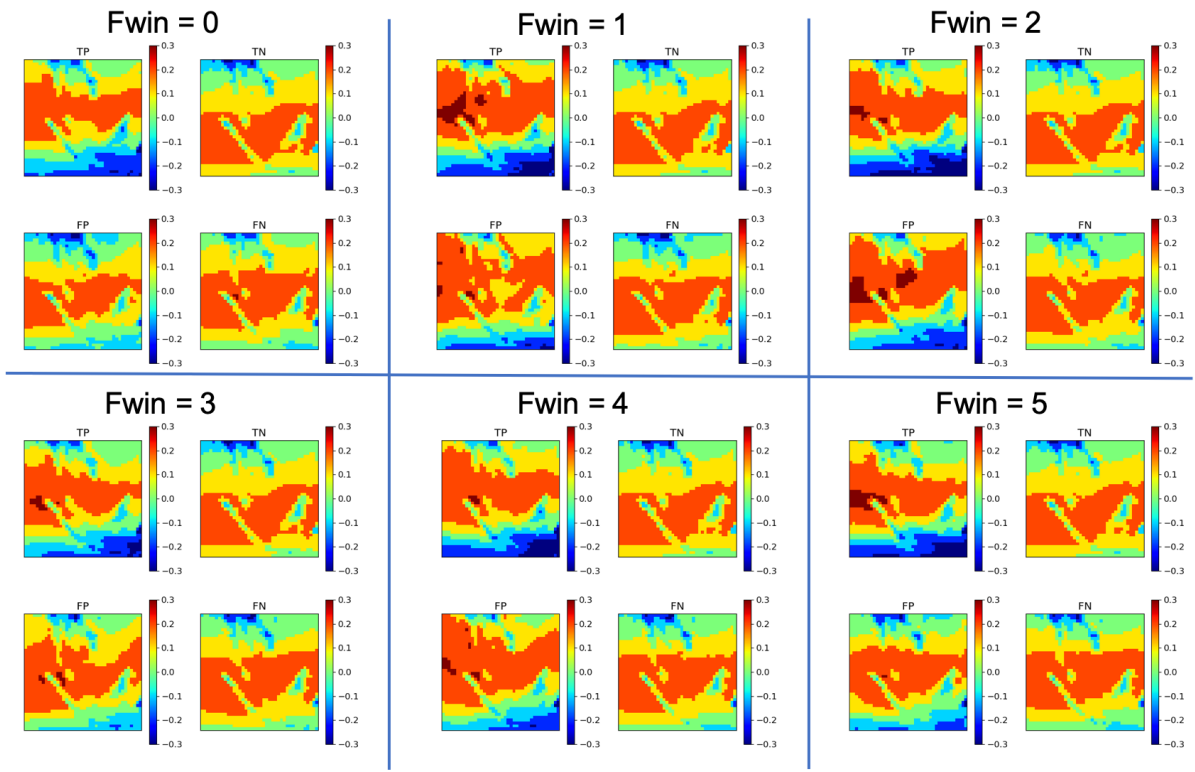

**Figure 9.** Average patterns of four outcomes in prediction, *i.e.*, TP = True Positive, TN = True Negative, FP = False Positive, and FN = False Negative derived by various machines trained using input features with different leading times (Fwin) in days. The patterns are blurred as in Fig. 5.

## 5. Summary

A deep learning-based framework has been developed and trained for forecasting the occurrence of severe haze events in Southeast Asia, specifically in Singapore. To proceed the supervised training, observed surface visibility from 1982 to 2016 have been used to label the occurrence of severe hazes, while geographic maps covering Southeast Asia of up to 18 meteorological and hydrological variables have been used as input features.

The machine has achieved a reasonable performance measured by validation accuracy, error, alongside many other metrics including precision, F1 score etc. One obvious shortcoming of the current machine is its producing relatively high number of missing forecasts (or false negatives). In-depth analysis looking through the similarity and difference between falsely predicted and observed patterns could benefit the improvement of the machine.

This practice can benefit advancing knowledge about, *e.g.*, the favorite environmental conditions of such an extreme event. The scale and depth of data analyses involved in the development of the deep learning framework would warrant a better generalization of conclusion from this study over a traditional case study.

**Acknowledgements**
This study is supported by L'Agence National de la Recherche (ANR) of France under "Programme d'Investissements d'Avenir" (ANR-18-MPGA-003 EUROACE), and by the National Research Foundation of Singapore through Singapore-MIT Alliance for Research and Technology (SMART). The author thanks A. Y.-M. Tonks for testing a prototype of the machine, A. Chulakadabbe and H.-H. Lee for experimenting various machine learning models that benefited this later deep learning effort, and R. Bar-Or and B. Grandey for discussions on data usage. The majority of the computations have been conducted using GPU clusters of French Grand equipment national de calcul intensif (GENCI) (Project 101056) and the CNRS mesocenter of computing of CALMIP (Project p18025).